\documentclass[runningheads]{llncs}

\usepackage{latexsym}
\usepackage{pstricks,pst-node,pst-text,pst-3d}

\newenvironment{08example}{\begin{example}}{\end{example}}
\newcommand{\reach}{\stackrel{\tiny reach}{\longrightarrow}}
\newcommand{\pow}[1]{2^{#1}}

\newcommand{\lcons}{\mbox{\tt '.'}}

\begin{document}
\setcounter{page}{63}
\title{More Precise Yet Efficient Type Inference for Logic Programs}
\titlerunning{More Precise Yet Efficient Type Inference for Logic Programs}
\author{Claudio Vaucheret \and Francisco Bueno}
\authorrunning{C. Vaucheret, F. Bueno}
\institute{Technical University of Madrid (UPM), Spain\\
           \email{\{claudio,bueno\}@fi.upm.es}}

\maketitle

\addtocounter{footnote}{1}
\footnotetext{In Alexandre Tessier (Ed), proceedings of the 12th International Workshop on Logic Programming Environments (WLPE 2002), July 2002, Copenhagen, Denmark.\\Proceedings of WLPE 2002: \texttt{http://xxx.lanl.gov/html/cs/0207052} (CoRR)}

\begin{abstract}
Type analyses of logic programs which aim at inferring the types of
the program being analyzed are presented in a unified abstract
interpretation-based framework. This covers most classical abstract
interpretation-based type analyzers for logic programs, built on
either top-down or bottom-up 
interpretation of the program. In this setting, we discuss the
widening operator, arguably a crucial one. We present a new widening
which is more precise than those previously proposed. Practical
results with our analysis domain are also presented, showing that it
also allows for efficient analysis.
\end{abstract}

\section{Introduction}

In type analyses, the widening operation has much influence in the
results. If the widening is too aggressive in making approximations
then the analysis results may be too imprecise. On the other hand,
if it is not sufficiently aggressive then the analysis may become
too inefficient. 

Widening operators are aimed at identifying the recursive structure
of the types being inferred. All widenings already proposed
in the literature are based on locating type nodes with the same
functors, which are possible sources of recursion. However, they
disregard whether such nodes come in fact from a recursive structure
in the program or not. This may originate an unnecessary loss of
precision, since the widening result may then impose a recursive
structure on the resulting type in argument positions where the
concrete program is in fact not recursive. We propose a widening
operator to try to remedy this problem. 

We present our widening operator for regular type inference in an
analysis framework based on abstract interpretation of the program.
In order for the paper to be self contained, we first revisit
regular types (Section~\ref{sec:regtypes}) and, in particular,
deterministic ones. We focus on deterministic types for ease of
presentation; however, there is nothing in our widening which prevents
it to be applicable also to non-deterministic types. The abstract
interpretation framework is set up in
Section~\ref{sec:domain}. Section~\ref{sec:widenings} 
reviews previous widenings in the literature, and
Section~\ref{sec:grammarwithnames} presents ours. In
Section~\ref{sec:results} experimental results are presented, and
Section~\ref{sec:conclusions} concludes and discusses future work. 

\section{Regular Types}
\label{sec:regtypes}

A {\em regular type}~\cite{Dart-Zobel} is a type representing a class
of terms that can be described by a regular term grammar.
A \emph{regular term grammar}, or grammar for short, describes a set of
finite terms constructed from a finite alphabet ${\cal F}$ of
\emph{ranked function symbols} or \emph{functors}.
A grammar $G = (S,{\cal T},{\cal F},{\cal R})$ consists of a
set of non-terminal symbols ${\cal T}$, one distinguished
symbol $S\in{\cal T}$, and a finite set ${\cal R}$ of
productions $T\longrightarrow rhs$, where $T \in {\cal T}$ is a
non-terminal and the right hand side $rhs$ is either a non-terminal or
a term $f(T_1,\ldots,T_n)$ constructed from an $n$-ary function symbol
$f \in {\cal F}$ and $n$ non-terminals. 

The non-terminals ${\cal T}$ are {\em types} describing (ground) terms
built from the functors in ${\cal F}$. The concretization $\gamma(T)$
of a non-terminal $T$ is the set of terms derivable from its productions,
that is, 
\begin{eqnarray*}
\gamma(T) & = & \bigcup_{(T \longrightarrow rhs) \in {\cal R}}
      \gamma(rhs) \\
\gamma(f(T_1,\ldots,T_n)) & = & \{ f(t_1,\ldots,t_n) ~|~
                                     t_i \in \gamma(T_i) \}
\end{eqnarray*}

The types of interest are each defined by one grammar: each $T_i$ is
defined by a grammar $(T_i,{\cal T}_i,{\cal F},{\cal R}_i)$, so that for
any two types of interest $T_1$ and $T_2$ on ${\cal F}$, 
${\cal T}_1\cap{\cal T}_2=\emptyset$. Sometimes, we will be interested
in types defined by non-terminals of a grammar 
$(T,{\cal T},{\cal F},{\cal R})$ other than the distinguished
non-terminal $T$. This is formalized by defining a type $T_i\in{\cal T}$
as the grammar  
\begin{equation}\label{eq:restrict}
(T_i,\{ T\in {\cal T} ~|~ T_i \reach^*_{\cal R} T \},
               {\cal F},\{ (T \longrightarrow rhs)\in {\cal R}
                        ~|~ T_i \reach^*_{\cal R} T \})
\end{equation}
where all the non-terminals are renamed apart, $\reach^*_{\cal R}$ is
the reflexive and transitive closure of $\reach_{\cal R}$ and
 $T_i \reach_{\cal R} T_j
   \mbox{ iff } T_i \longrightarrow_{\cal R} T_j 
   \mbox{ or }  T_i \longrightarrow_{\cal R} f(\ldots,T_j,\ldots)$.

A grammar is in \emph{normal form} if none of the right hand sides are
non-terminals.  A particular class of grammars are deterministic ones. 
A grammar is \emph{deterministic} if it is in normal form
and for each non-terminal $T$ the function symbols are all distinct in
the right hand sides of the productions for $T$.  

Deterministic grammars are less expressive than non-deterministic ones.
Deterministic grammars can only express sets of terms which are
\emph{tuple-distributive}; informally speaking, which are ``closed
under exchange of arguments''. I.e., if the set contains two terms of
the same functor, then it also contains terms with the same principal
functor obtained by exchanging subterms of the previous two terms in
the same argument positions. Basically, no dependencies between
arguments of a term can be expressed with deterministic grammars.

\begin{08example} 
Consider the type $T$ denoting the set $\{f(a,b),f(c,d)\}$, which is
non-deterministic,
$$\begin{array}{lllllllllll}
  T & \longrightarrow & f(A,B) && A & \longrightarrow & a && C & \longrightarrow & c \\ 
  T & \longrightarrow & f(C,D) && B & \longrightarrow & b && D & \longrightarrow & d    
  \end{array}$$
A deterministic type $T'$ with a concretization which included
$\gamma(T)$ would also have to include $\{f(c,b),f(a,d)\}$, that is,
$$\begin{array}{lllllllllll}
  T' & \longrightarrow & f(AC,BD) && AC & \longrightarrow & a && BD & \longrightarrow & b \\
     &                 &          && AC & \longrightarrow & c && BD & \longrightarrow & d
  \end{array}$$
To facilitate the presentation non-terminals with a single production
will often be ``inlined'' and multiple right hand sides combined so
that $T$ above will be written 
$T \longrightarrow f(a,b) \ | \ f(c,d)$ and $T'$ as
$$\begin{array}{lllllllllll}
  T' & \longrightarrow & f(AC,BD) && AC & \longrightarrow & a \  | \ c &&
  BD & \longrightarrow & b \  | \ d
  \end{array}$$
\vspace{-2\baselineskip}
\end{08example}

To be able to describe terms containing numbers and variables we
introduce two distinguished symbols \textbf{num} and \textbf{any},
plus an additional $\bot$. The concretization of \textbf{num} is the
set of all numbers, the concretization of \textbf{any} is the set
of all terms (including variables), and the concretization of $\bot$ is
the empty set of terms. These symbols are non-terminals but they are
considered terminals to the effect of regarding a grammar as
deterministic. 

Let $\mathcal{G}$ be the set of all grammars, if $T_1$, $T_2$ belong to
$\mathcal{G}$, the relation $T_1 \equiv T_2 \Leftrightarrow \gamma(T_1) =
\gamma(T_2)$ is an equivalence relation. The quotient set
$\mathcal{G}/\equiv$ is a complete lattice with top element \textbf{any}
and bottom element $\bot$ based on the relation of {\em containment}, or
type {\em inclusion}: for every 
$\overline{T_1},\overline{T_2} \in \mathcal{G}/\equiv$,
$\overline{T_1} \sqsubseteq \overline{T_2}
\Leftrightarrow \gamma(T_1)\subseteq\gamma(T_2)$.
% where
% and $T_i$ is the representative of the equivalence class $\overline{T_i}$
%
% Non-terminals, and therefore,
% grammars are ordered by the relation of {\em containment}, or type {\em
%   inclusion}, $(T_1 \sqsubseteq T_2) \Leftrightarrow
% \gamma(T_1)\subseteq\gamma(T_2)$.  Based on this, the set of all grammars
% $\mathcal{G}$ is a complete lattice with top element \textbf{any} and
% bottom element $\bot$. 
%
We will denote $\overline{T_i}$ simply by $T_i$.

The least upper bound is given by type {\em union},
$({T_1} \sqcup {T_2})$, and the greatest lower bound by 
type {\em intersection}, 
$({T_1} \sqcap {T_2})$~\cite{Dart-Zobel}. It can be shown that
intersection describes term unification:
\[ t_1^*\subseteq\gamma(T_1)\wedge t_2^*\subseteq\gamma(T_2)\wedge 
   t_1\theta=t_2\theta \Rightarrow 
   (t_1\theta)^*\subseteq\gamma(T_1 \sqcap T_2)
\]
where $t^*$ denotes the set of ground terms which are instances of the
term $t$.

\section{Abstract Domain for Type Inference}
\label{sec:domain}

In an abstract interpretation-based type analysis, a type is
used as an abstract description of a set of terms. Given variables of
interest $\{x_1,\ldots,x_n\}$, any substitution
$\theta = \{x_1 \leftarrow t_1,\ldots,x_n \leftarrow t_n\}$
can be approximated by an {\em abstract substitution} 
$\{x_1 \leftarrow T_{x_1},\ldots,x_n \leftarrow T_{x_n}\}$
where $t_i \in \gamma(T_{x_i})$ and each type $T_{x_i} \in \mathcal{G/\equiv}$.
We will write abstract substitutions as tuples
$\langle T_1,\ldots,T_n \rangle$, and sometimes also abbreviate a
tuple simply as $T^n$.

Concretization is lifted up to abstract substitutions
straightforwardly, 
$$\begin{array}{rcl}
\gamma(\langle T_1,\ldots,T_n \rangle) & = & 
\{~\{x_1 \leftarrow t_1,\ldots,x_n \leftarrow t_n\}~|~t_i \in \gamma(T_i)~\}
\end{array}$$
as well as the equivalence relation $\equiv$. Additionally,
 we consider a distinguished abstract
substitution {\boldmath$\bot$} as a representative of any 
$\langle T_1,\ldots,T_n\rangle$ such that there is $T_i=\bot$.
Of course, $\gamma(${\boldmath$\bot$}$)=\emptyset$.

An ordering on the domain is obtained as the natural element-wise
extension of the ordering on types:
$$\begin{array}{rcl}
 \begin{array}[b]{rcl}
    \mbox{\boldmath $\bot$} & \sqsubseteq & T^n \\
    \langle T_1,\ldots,T_n \rangle & \not \sqsubseteq & \mbox{\boldmath $\bot$} \\
    \langle T_1,\ldots,T_n \rangle & \sqsubseteq & \langle T'_1,\ldots,T'_n \rangle \\
 \end{array} & \Longleftrightarrow & \forall_{1 \leq i \leq n}T_i
    \sqsubseteq T'_i
%%  \\
%% T^n  \equiv  T'^n & \Longleftrightarrow & T^n \sqsubseteq T'^n \wedge
%% T'^n \sqsubseteq T^n
\end{array}$$
The domain is a lattice with bottom element {\boldmath $\bot$} and top element
$\langle T_1,\ldots,T_n \rangle$ such that $T_1=\ldots=T_n=\textbf{any}$.
The greatest lower bound and least upper bound domain operations are
lifted also element-wise, as follows,
$$\begin{array}{rcl}
\mbox{\boldmath $\bot$} \sqcup T^n  =  T^n \sqcup \mbox{\boldmath $\bot$} & = & T^n \\
\langle T_1,\ldots,T_n \rangle \sqcup \langle T'_1,\ldots,T'_n \rangle
& = & \langle T_1 \sqcup T'_1 ,\ldots,T_n \sqcup T'_n \rangle \\
\mbox{\boldmath $\bot$} ~ \sqcap ~ T^n  =  T^n ~ \sqcap ~ \mbox{\boldmath $\bot$} 
& = & \mbox{\boldmath $\bot$} \\
\langle T_1,\ldots,T_n \rangle ~ \sqcap ~ \langle T'_1,\ldots,T'_n \rangle
& = & \langle T_1 \sqcap T'_1 ,\ldots,T_n \sqcap T'_n \rangle
%% \bot \bigtriangledown  \langle T'_1,\ldots,T'_n \rangle 
%% & = & \langle \bot \bigtriangledown T'_1,\ldots,\bot \bigtriangledown T'_n \rangle
%% \\
%% \langle T_1,\ldots,T_n \rangle \bigtriangledown \langle T'_1,\ldots,T'_n \rangle
%% & = & \langle T_1 \bigtriangledown T'_1 ,\ldots,T_n \bigtriangledown T'_n \rangle
\end{array}$$

Using the adjoint $\alpha$ of $\gamma$ as abstraction function, it can
be shown that $(\pow{\Theta},\alpha,\Omega,\gamma)$ is a Galois
insertion, where $\Theta$ is the domain of concrete % substitutions 
and $\Omega$ that of abstract substitutions. 
The following abstract unification operator can be shown to approximate the
concrete one. Let $x=t$ be a concrete unification equation, with $x$ a
variable,$t$ any term, and $T^n$ the current abstract substitution, and
let $y_j$, $j=1,\ldots,m$ be the variables of $t$,
the new abstract substitution is:
\begin{equation}\label{eq:amgu}
amgu(T^n,x=t)=
T^n[T_x/T'_x,T_{y_1}/T'_{y_1},\ldots,T_{y_m}/T'_{y_m}]
\end{equation}
with each $T$ replaced by $T'$ in the
tuple, $T'_x = T_x\sqcap t\mu$, 
$\mu=\{y_1\leftarrow T_{y_1},\ldots,y_m\leftarrow T_{y_m}\}$, and 
$solve(t,T'_x) = \{ y_1=T'_{y_1},\ldots,y_m=T'_{y_m} \}$, a set of
equations that define the 
types of the variables of a term $t\in\gamma(T'_x)$, obtained as:
%\marginpar{revise solve}
$$solve(t,T)=\left\{
%\mbox{\begin{eqnarray*}
\begin{array}{lll}
\{t=T\}             & \mbox{if} & t \mbox{ is a variable} \\
{\displaystyle \bigcup_{T \longrightarrow f(T_1,\ldots,T_n)}\ \bigcup_{i=1,\ldots,n}}
solve(t_i,T_i)      & \mbox{if} & t \mbox{ is } f(t_1,\ldots,t_n)
\end{array}
%\end{eqnarray*}}
\right.$$

In this abstract interpretation-based setting, analysis with a
monotonic semantic function can be easily
shown correct. However, it is not guaranteed to terminate, since
$\Omega$ has infinite ascending chains. To guarantee termination, a
widening operator is required. 

\begin{08example}
\label{ex:listoflists}
%% Consider the following program which defines the regular type lists of lists of
%% numbers: 
The following program defines the type lists of lists of numbers: 
\begin{verbatim}
list_of_lists([]).                  num_list([]).        
list_of_lists([L|Ls]):-             num_list([N|Xs]):-   
        num_list(L),                        number(N),  
        list_of_lists(Ls).                  num_list(Xs).
\end{verbatim}
For the argument of \texttt{num\_list}, without a widening operator,
an analysis would obtain the following first three approximations:  
$$\begin{array}{rclccrclccrclccrcl}
T_0 & \longrightarrow & [] & & &
T_1 & \longrightarrow & [] \ | \ .(\mathbf{num},T_0) & & &
T_2 & \longrightarrow & [] \ | \ .(\mathbf{num},T_1)
\end{array}$$
where each $T_i$ represents a list of $i$ numbers. Analysis will
never terminate, since it would keep on obtaining a new type
representing a list with one more number. A widening operator would be
required that over-approximates some type $T_l$ to something like
$T_l \longrightarrow [] \ | \ .(\mathbf{num},T_l)$,
%\centerline{$T_l \longrightarrow [] \ | \ .(\mathbf{num},T_l)$}
%\noindent
which is the expected type, and allows termination of the analysis.
\end{08example}

\section{Widenings}
\label{sec:widenings}

%% A widening is required to guarantee that an analysis
%% terminates when the abstract domain has infinite ascending chains as
%% is the case of regular types. 
%% We now revisit widening operators for types proposed in the literature.

%% \begin{definition}[widening operator] \label{def:widening}
%% An operator $\bigtriangledown$ is a widening iff it is:
%%   \begin{description}
%%   \item[correct] $a_1 \sqsubseteq a_1 \bigtriangledown a_1 \wedge
%%     a_2 \sqsubseteq a_1 \bigtriangledown a_2$
%%   \item[stationary] for every increasing chain $a_0 \sqsubset a_1
%%     \sqsubset a_2 \sqsubset \ldots$, the chain $b_0 = a_0, b_1 = b_0
%%     \bigtriangledown a_1, \ldots ,b_{i+1} = b_i \bigtriangledown
%%     a_{i+1}, \ldots$ is not strictly increasing for $\sqsubseteq$,
%%     i.e., there is an $n$ such that $b_n=b_{n+1}$.
%%   \end{description}
%% \end{definition}
%% 
%% For the purposes of termination of the analysis it is enough to
%% require that the widening is stationary for the ascending chains that
%% may occur during analysis of a program.

\paragraph{Functor Widening} \label{sec:widen-de-funct}
This is probably the simplest widening operator which still
keeps information from the recursive structure of the program that
``produces'' the corresponding terms. The idea behind it is to create
a type and a production for each functor symbol in the original type.
All arguments of the function symbols are replaced with the new types~\cite{mildnerthesis}. 

\begin{08example} \label{ex:widen-de-funct}
Consider predicate \texttt{list\_of\_lists} of
Example~\ref{sec:domain}.\ref{ex:listoflists}, its argument should 
ideally have the following type:
%% \begin{eqnarray*}
%%     T_{ll} & \longrightarrow & [] \  | \ .(T_l,T_{ll})  \\
%%     T_l & \longrightarrow & [] \  | \ .(\mathbf{num},T_l) 
%% \end{eqnarray*}
$\begin{array}{lclcclcl}
    T_{ll} & \longrightarrow & [] \  | \ .(T_l,T_{ll}) & & &
    T_l & \longrightarrow & [] \  | \ .(\mathbf{num},T_l) 
\end{array}$
but the functor widening will yield:
$T \longrightarrow [] \ | \ \mathbf{num} \ | \ .(T,T) $.
%% \centerline{$T \longrightarrow [] \ | \ \mathbf{num} \ | \ .(T,T) $}
%\vspace{-2.5\baselineskip}
\end{08example}

\paragraph{Type Jungle Widening} \label{sec:widening-jungla-de}
A type jungle is a grammar where each functor always has the same
arguments. It was originally proposed as a finite type
domain~\cite{lindgrenmildner} , 
since in a domain where all grammars are of the type jungle class
all ascending chains are finite. However, it can be used as a
subdomain to provide a widening %operator.

\begin{08example}
Applying this widening to the previous type $T_{ll}$, the following
will be obtained:
%% \begin{eqnarray*}
%%   T & \longrightarrow & [] \ | \ .(T_1,T)  \\
%%   T_1 & \longrightarrow & [] \ | \ \mathbf{num} \ | \ .(T_1,T)
%% \end{eqnarray*}
\vspace{-\baselineskip}
$$\begin{array}{lclcclcl}
  T & \longrightarrow & [] \ | \ .(T_1,T) & & &
  T_1 & \longrightarrow & [] \ | \ \mathbf{num} \ | \ .(T_1,T)
\end{array}$$
%\vspace{-2.5\baselineskip}
\end{08example}
 
Note that this widening is strictly more precise than the functor
widening. In the example, the new type captures the upper level of
lists, but it loses precision when describing the type of the list
elements. This is due to the restriction of forcing functors to always
have the same arguments. 

\paragraph{Shortening} \label{sec:shortening}
A grammar can be seen as a graph where the nodes correspond to the
non-terminals (or-nodes) and to the right hand sides of productions
(and-nodes), and the edges correspond to the production relation or
the relation between a functor and its arguments in a right hand side
of a production.
Given an or-node, its {\em principal functors} are the functors
appearing in its children nodes.

\begin{08example}
The type $T_{ll}$ of the previous examples can be seen as the graph:

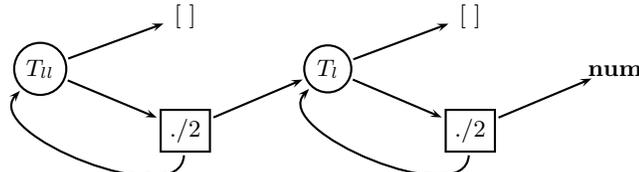
\begin{small}
%\vspace{-0.5\baselineskip}
\begin{center}
$
\psmatrix[colsep=9ex,rowsep=1ex]
&                & \mbox{ [ ] }       &             & \mbox{ [ ] } \\
& [mnode=circle] T_{ll} &             & [mnode=circle] T_l & & \mathbf{num} \\
&                & \psframebox{ ./2 } &             & \psframebox{ ./2 } \\
\endpsmatrix 
\psset{nodesep=1pt,arrows=->}
\ncline{2,2}{1,3}
\ncline{2,2}{3,3}
\ncline{3,3}{2,4}
\ncline{2,4}{1,5}
\ncline{2,4}{3,5}
\ncline{3,5}{2,6}
\ncarc[arcangle=110]{3,3}{2,2} 
\ncarc[arcangle=110]{3,5}{2,4} 
$
\end{center}
\end{small}
\end{08example}

Gallagher and de Waal~\cite{gallagher-types-iclp94} defined a widening
which avoids having two or-nodes, which have the same principal functors,
connected by a path. If two such nodes exist, they are replaced
by their least upper bound.

\begin{08example}
In the above example graph, nodes $T_{ll}$ and $T_l$ have the same
principal functors ([] and .) so that they are replaced, yielding:
%% \begin{eqnarray*}
%%   T & \longrightarrow & [] \ | \ .(T_1,T)  \\
%%   T_1 & \longrightarrow & [] \ | \ \mathbf{num} \ | \ .(\mathbf{num},T)
%% \end{eqnarray*}
$$\begin{array}{lclcclcl}
  T & \longrightarrow & [] \ | \ .(T_1,T) & & &
  T_1 & \longrightarrow & [] \ | \ \mathbf{num} \ | \ .(\mathbf{num},T)
\end{array}$$
%\vspace{-2.5\baselineskip}
\end{08example}

Note the precision improvement with respect to the result in the
previous example. Note also that still the result is imprecise.

\paragraph{Restricted Shortening} \label{sec:short-restr}
Saglam and Gallagher~\cite{Saglam-Gallagher-94} propose a more precise
variant of the previous widening. Shortening is restricted so that two
or-nodes $T$ and $T'$ which are connected by a path from $T$ to $T'$
and have the same principal functors are replaced only if 
$T'\sqsubseteq T$. If this is the case, only $T'$ needs be replaced,
since the least upper bound is $T$.  

\begin{08example} \label{ex:short-restr}
Continuing previous examples, since nodes $T_{ll}$ and $T_l$ have
the same principal functors but $T_l \not \sqsubseteq T_{ll}$, 
the widening operation will make no change. In this case, the most
precise type is achieved.
\end{08example}

Note, however, that restricted shortening does not guarantee
termination in general (and thus, it is not, strictly speaking, a
widening). There are cases in which analysis may not 
terminate using only this widening operator~\cite{mildnerthesis}.

\paragraph{Depth Widening}
Janssens and Bruynooghe~\cite{janss92} proposed a type analysis in
which the widening effect is achieved by a ``pruning'' of the type
depth up to a certain bound. A parameter k establishes the maximum
number of occurrences of a functor in-depth in a type. The idea is
similar to the well-known depth-k abstraction for term structure
analysis. The resulting type analysis uses normal restricted type
graphs, which are basically deterministic types satisfying the depth
limit. Obviously, the precision % of this analysis
depends on the value of the parameter k. 

\begin{08example}
The widening of our previous type $T_{ll}$ with k=1 will yield the
same result than the functor widening
(Example~\ref{sec:widenings}.\ref{ex:widen-de-funct}), whereas with
k=2 will yield the same result as restricted shortening
(Example~\ref{sec:widenings}.\ref{ex:short-restr}). 
\vspace{-\baselineskip}
\end{08example}

\paragraph{Topological Clash Widening}
Van Hentenryck et al.~\cite{VanHentenryckCortesiLeCharlier95} proposed
the first widening operator that takes  into account two consecutive
approximations to the type being inferred. After merging the two
---i.e., calculating their least upper bound, the result is compared
with the previous approximation to try to ``guess'' where the type is
growing. This is done by locating {\em topological clashes}: functors
that differ or appear at different depth in each type graph. 
%\marginpar{revise!}
The clashes are resolved by replacing them with the recently
calculated least upper bound. 

\begin{08example}
\label{ex:sorted}
Consider the program:
\begin{verbatim}
sorted([]).
sorted([_X]).
sorted([X,Y|L]):- X =< Y, sorted([Y|L]).
\end{verbatim}
and the moment during analysis when the final widening is performed.
The resulting type for the argument of \texttt{sorted/1}
is the one on the left below for the first two clauses, and the one on
the right for the last one:
$$\begin{array}{lclcclcl}
    T_0 & \longrightarrow & [] \  | \ .(\mathbf{any},[]) & & &
    T_1 & \longrightarrow & .(\mathbf{num},.(\mathbf{num},T_l))  \\
        &                 &                               & & &
    T_l & \longrightarrow & [] \  | \ .(\mathbf{num},T_l) 
\end{array}$$
Their least upper bound is $T_u$ on the left below, which
exhibits a clash with $T_0$ in the second argument of functor $./2$. 
Thus, the result of widening is $T_s$: % on the right:
$$\begin{array}{lclcclcl}
  T_u & \longrightarrow & [] \ | \ .(\mathbf{any},T_l) & & &
  T_s & \longrightarrow & [] \ | \ .(\mathbf{any},T_s)
\end{array}$$
%% \begin{eqnarray*}
%%   T_s & \longrightarrow & [] \ | \ .(\mathbf{any},T_s)
%% \end{eqnarray*}
%\vspace{-1.5\baselineskip}
\end{08example}

All widening operators are based on locating recursive
structures in the type definitions where there are nodes with the same
functors. This may originate an unnecessary loss of precision, since
the widening may impose a recursive structure on the
resulting type in argument positions where the concrete program is in
fact not recursive. In the following section we present a new widening
operator that tries to remedy this problem.

\section{Structural Type Widening}
\label{sec:grammarwithnames}

In this section we define an extended domain for type analysis which
incorporates a widening operator aimed at improving the precision of
the analysis. The domain is defined so as to keep track of information
on the program structure, so that recursion on the types produced by
the analysis is imposed by the widening operator only in the cases
where it corresponds to a recursive structure in the program being
analyzed. To this end, type names will be used.

A \emph{type name} is roughly a (distinguished) non-terminal that
represents a 
type produced during the analysis. Type names are created for each
variable in each argument of each variant of each program atom for
each predicate (note how this is different from, for example,
set-based analyses~\cite{set-based-popl}, where variants are not taken into
account).

Type names provide information on how types are being formed from
other types during analysis. This makes it possible to precisely
identify places where to impose recursion on the types: in a
subterm of the type which happens to refer to the name of that
type. To this end, type names contain 
references to the position of its constituent types. To determine
positions, selectors are used, as defined below.

\begin{definition}[selector]
    Define $t/s$, the subterm of a concrete term $t$ referenced by a
    \emph{selector} $s$, inductively as follows. The empty selector
    $\epsilon$ refers to the term $t$, that is, $t/\epsilon = t$. 
    If $t/s = t'$, $t'$ is a compound term 
    $f(t_1,\ldots,t_i,\ldots,t_n)$ (where $f$ is an $n$-ary function
    symbol) then $t/s\cdot(f.i) = t_i$, $1 \leq i \leq n$. 
\end{definition}

For every two selectors $s$, $p$, if $t/s=t'$ and if $t'/p$ exists
then $t/s\cdot p=t'/p$. The initial $\epsilon$ of a non-empty selector
will often be omitted, so $\epsilon\cdot p$ will be written simply as
$p$. 

We define a set of type names $\mathcal{N}$ such that
$\mathcal{N}\cap\mathcal{G}=\emptyset$ and a set
$\pow{\mathcal{N}\times\mathcal{G}}$ of relations 
$\mathcal{X}\in\pow{\mathcal{N}\times\mathcal{G}}$
between type names and types, of the form
$\mathcal{X}\subseteq\mathcal{N}\times\mathcal{G}$.

\begin{definition}[label]
  Let $\mathcal{X}$ a relation between type names and types. 
  Given a type name $N$, a \emph{label} of $N$ is a tuple
  $\langle s,N' \rangle$, where $s$ is a selector and $N'$ is a
  type name, iff $(N,T)\in\mathcal{X}$, $(N',T')\in\mathcal{X}$, and
  $T' \sqsubseteq T/s$.
%\marginpar{$T/s \sqsubseteq T'$ ??}
\end{definition}

Labels of a type name $N$ indicate subterms of the type $T$ defining
$N$ where other type names occur.

\begin{08example}
Let a relation $\mathcal{X}$ such that $\{ (A,T_1), (B,T_2) \}
\subseteq \mathcal{X}$, and let grammars 
$(T_1,{\cal T}_1,{\cal F},{\cal R}_1)$ and
$(T_2,{\cal T}_2,{\cal F},{\cal R}_2)$, such that
the only rule for $T_1$ is
$(T_1 \longrightarrow f(b))\in {\cal R}_1$ and
$(T_2 \longrightarrow g(c,T_3))\in {\cal R}_2$,
$(T_3 \longrightarrow b ~|~ f(b))\in {\cal R}_2$.
Consider a label $\langle (g.2),A \rangle$ of $B$.
We have that $T_1 \sqsubseteq T_2/(g.2) = T_3$.
\end{08example}
%\vspace{-0.5\baselineskip}

\begin{definition}[type descriptor]
  Let $\mathcal{G}$ a set of types (regular term grammars),
  $\mathcal{N}$ a set of type names, and
  $\mathcal{X}\subseteq\mathcal{N}\times\mathcal{G}$. A
  \emph{type descriptor} is a tuple $(N,E,T)$ where $N\in\mathcal{N}$,
  $T\in\mathcal{G}$, $(N,T)\in\mathcal{X}$, and $E$ is a set of labels
  of $N$. 
\end{definition}

In the new domain, type descriptors will be used instead of types. Let
$\mathcal{D}$ be the set of all type descriptors from given sets of
types $\mathcal{G}$ and of type names $\mathcal{N}$. Concretization is
defined as $\gamma((N,E,T))=\gamma(T)$. The domain ordering and
operations on $\mathcal{D}$ are the same as on $\mathcal{G}$ except
for type names. In this case, they have to take into account the
possible labels of the type name.

%\vspace*{\baselineskip} 
%\begin{nondefinition}[Inclusion]
\paragraph{Inclusion}
%\noindent{\bf Inclusion}
$(N_1,E_1,T_1) \sqsubseteq (N_2,E_2,T_2) \Leftrightarrow 
T_1 \sqsubseteq T_2 \wedge E_1 \subseteq E_2$.

\paragraph{Union}
%\noindent{\bf Union}
$(N,E,T) = (N_1,E_1,T_1) \sqcup (N_2,E_2,T_2) \Leftrightarrow 
T = T_1 \sqcup T_2 \wedge E = E_1 \cup E_2$.

\paragraph{Intersection}
%\noindent{\bf Intersection}
$(N,E,T) = (N_1,E_1,T_1) \sqcap (N_2,E_2,T_2) \Leftrightarrow 
T = T_1 \sqcap T_2 \wedge E = E_1 \cup E_2$.
%\end{nondefinition}
%\vspace*{\baselineskip} 

\paragraph{}
Again, we may be interested in types defined by non-terminals other
than the distinguished non-terminal $T$ of a grammar
$(T,{\cal T},{\cal F},{\cal R})$. A type descriptor $(N_i,E_i,T_i)$,
where $T_i\in{\cal T}$, is formally defined from $(N,E,T)$ as
follows: $T_i$ is the grammar of Equation~\ref{eq:restrict}, $N_i$ is
a new type name, and 
$$E_i = \{ \langle p,N' \rangle ~|~ \langle s\cdot p,N' \rangle \in E
                \wedge T/s = T_i \}.$$

Abstract substitutions for variables of interest $\{x_1,\ldots,x_n\}$
are now defined as tuples of the form
$\langle (N_1,E_1,T_{x_1}),\ldots,(N_n,E_n,T_{x_n})\rangle$. 
Concretization and the domain ordering and operations are lifted to
abstract substitutions element-wise, in the same way as in
Section~\ref{sec:domain}, including the widening operator defined
below. 
If now $\Omega$ is the domain of type descriptors, it can be shown
that $(\pow{\Theta},\alpha,\Omega,\gamma)$ is a Galois insertion,
where $\alpha$ is the adjoin of $\gamma$. Abstract unification is
defined as in Equation~\ref{eq:amgu}, but using type descriptors
instead of types (and preserving all type names in the ``input''
abstract substitution $T^n$ to $amgu$).

\begin{definition}[structural widening]
The widening between an approximation $T_2$ to type name $N$ and a
previous approximation $T_1$ to $N$ is
$(N,E_1,T_1) \bigtriangledown (N,E_2,T_2) = (N,E_1 \cup E_2,T)$, such
that $T$ is defined by $(T,{\cal T},{\cal F},{\cal R})$ where 
${\cal T}=\{ T_i ~|~ T \longrightarrow^*_{\cal R} T_i \}$,
and ${\cal R}$ is obtained by the following algorithm:
\end{definition}

\begin{tabbing}
$T'$ := $T_1 \sqcup T_2$ 
        defined by $(T',{\cal T}',{\cal F},{\cal R}')$ \\
$\mathcal{S}$ := $\{ s ~|~ (s,N) \in E_1\cup E_2 \}$ \\
$Seen := \emptyset$ \\
\texttt{for }\=\texttt{each} 
        $(T' \longrightarrow f(A_1,\ldots,A_n))\in {\cal R}'$ 
        \texttt{add to ${\cal R}$ production} \\ 
 \> $T \longrightarrow 
       f(\texttt{widen}(A_1,{\cal R}',(f.1)),\ldots,
         \texttt{widen}(A_n,{\cal R}',(f.n)))$ \\
\\
\texttt{widen}(\=$N,{\cal R}',Sel) :$  \\
\> \texttt{if} $N$ = \textbf{any} \texttt{return} \textbf{any} \\
\> \texttt{if} $\exists M \langle N,M \rangle \in Seen$
  \texttt{return} $M$ \\
\> \texttt{let} $M$ \texttt{a new non-terminal} \\
\> $Seen$ \texttt{:=} $Seen \cup \{ \langle N,M \rangle \}$ \\
\> \texttt{for }\=\texttt{each} 
      $(N \longrightarrow f(A_1,\ldots,A_n))\in {\cal R}'$ 
      \texttt{add to ${\cal R}$ production} \\ 
\> \> $M \longrightarrow 
       f(\texttt{widen}(A_1,{\cal R}',Sel\cdot(f.1)),\ldots,
         \texttt{widen}(A_n,{\cal R}',Sel\cdot(f.n)))$ \\
\> \texttt{if} \= Sel $\in \mathcal{S}$ \texttt{then} \\
\> \> \texttt{add to ${\cal R}$ production} $M \longrightarrow T$ \\
\> \texttt{return} $M$
\end{tabbing} 

Structural widening basically identifies subterms of the new type
$T_1 \sqcup T_2$ where a reference to the type $N$ being widened
appear, and makes this ``self-reference'' explicit in the definition
of the new type. 
%
% Note that the result of the widening, $T$ is a correct over-aproximation of
% $T_2$ since each subterm where a reference to the type $N$ appear is
% equivalent to a previous approximation, which is in turn included in $T_1
% \sqcup T_2$, therefore $T_2 \sqsubseteq (T_1 \sqcup T_2) \sqsubseteq T$ .
%
Note that the widening operation starts with the least upper bound and,
basically, adds new grammar rules to that type. Therefore, the result
is always a correct approximation of such an upper bound. This
justifies its correctness.
Moreover, this approach based on type names is potentially more precise
than any of the  previous widening operators discussed, as the
following examples show: 

\begin{08example}
Consider program \texttt{sorted} in Example~\ref{sec:widenings}.\ref{ex:sorted}.
A top-down analysis with topological clash was roughly described there.
Let us now look at analysis using restricted shortening. The resulting
type happens to be the same one.

Analysis of program atom \texttt{sorted([Y|L])} approximates variable
\texttt{Y} always as $\mathbf{num}$, both in the calls and in the
successes. The first two success approximations for variable
\texttt{L} are $[]$ and $.(\mathbf{num},[])$. Their lub (and 
widening) is:
\begin{eqnarray*}
  T_1 & \longrightarrow & [] \ | \ .(\mathbf{num},[])  
\end{eqnarray*}
The next approximation to the type of \texttt{L} is
$.(\mathbf{num},T_1)$. Its lub with $T_1$ is 
$T_2 \longrightarrow [] ~|~ .(\mathbf{num},T_1)$, and since $T_2$ and
$T_1$ have the same functors, and $T_1$ is included in $T_2$, the
widening of $T_2$ is: 
\begin{eqnarray*}
T_3 \longrightarrow [] \ | \ .(\mathbf{num},T_3)  
\end{eqnarray*}
i.e., list of numbers. The next approximation to the type of \texttt{L} 
is $.(\mathbf{num},T_3)$ (i.e., a list with at least one number). It
is included in $T_3$, so fixpoint is reached.

The success of principal goal \texttt{sorted(X)} is approximated
after analyzing the two non-recursive clauses by
$T_4 \longrightarrow [] \ | \ .(\mathbf{any},[])$. Analysis of the
third clause yields $.(\mathbf{num},.(\mathbf{num},T_3))$. Its lub
with $T_4$ is $T_5 \longrightarrow [] \ | \ .(\mathbf{any},T_3)$. The
widening of $T_5$ finds that $T_5$ and $T_3$ have the same functors and
$T_3 \sqsubseteq T_5$, since \textbf{num} $\sqsubseteq$ \textbf{any}.
Thus, the result of widening is:
\begin{eqnarray*}
T_6 \longrightarrow [] \ | \ .(\mathbf{any},T_6)
\end{eqnarray*}
i.e., list of terms. This is the final result after one more 
iteration. Note that the information about successes where the tail of  
lists of length greater than one is a list of numbers is lost.

Let us now consider structural widening. Analysis of atom
\texttt{sorted([Y|L])} always approximates the type of \texttt{Y} by
$(N_{13},\emptyset,\mathbf{num})$. For variable \texttt{L} the two first
approximations are $(N_{14},\emptyset,[])$ and
$(N_{14},E_{14},.(\mathbf{num},[]))$, where the set of labels is
$E_{14} = \{\ (\lcons.1,N_{13}),\ (\lcons.2,N_{14})\ \}$. 
The result of widening is $(N_{14},E_{14},T_1)$ where $T_1$ is defined
as: 
\begin{eqnarray*}
  T_1 & \longrightarrow & [] \ | \ .(\mathbf{num},T_1)
\end{eqnarray*}
i.e., list of numbers. This is the final result after one more
iteration. 

The success of principal goal \texttt{sorted(X)} is approximated
after analyzing the two non-recursive clauses by $(N_3,\emptyset,T_2)$
where $T_2 \longrightarrow [] \ | \ .(\mathbf{any},[])$. Analysis of
the third clause yields
$(N_{3},E_3,.(\mathbf{num},.(\mathbf{num},T_1)))$, where
\begin{eqnarray*}
  E_3 & = & \{\ (\lcons.2\cdot \lcons.1,N_{13}),\
                (\lcons.2\cdot \lcons.2,N_{14})\ \}
\end{eqnarray*}
%\marginpar{$(\lcons.1,N_6)$, $(\lcons.2,N_2)$ ??}
Its widening with the previous approximation $T_2$ is $(N_3,E_3,T_3)$,
where 
\begin{eqnarray*}
  T_3    & \longrightarrow & [] \ | \ .(\mathbf{any},T_1)
%%   E(N_3) & = & \{ (\lcons.1,N_6) , (\lcons.2,N_2) , 
%%                   (\lcons.2\cdot \lcons.1,N_{13}) , 
%%                   (\lcons.2\cdot \lcons.2,N_{14}) \}
\end{eqnarray*}
which amounts to their lub, since the widening operator does not
produce any change, because $N_3$ is not among its own labels.
Therefore, the final result, after one more iteration, is $T_3$,
where indeed lists of length greater than one have a tail which is 
a list of numbers. 
\end{08example}

However, structural widening does not guarantee termination. It is
effective as long as the new approximation is built from the previous
approximation of the type being inferred. This case is identified, in
essence, by locating a reference to the type name of the previous
approximation within the definition of the new one. However, there are
contrived
cases in which a type is constructed during analysis which loses the
reference to the previous approximation. In these cases, a more
restrictive widening has to be applied to guarantee termination.

\begin{08example}
\label{ex:nontermination}
Consider the program:
\begin{verbatim}
main:- p(a).     p(a).                    q(a,f(a)).
                 p(X):- q(X,Y), p(Y).     q(f(Z),f(L)):- q(Z,L).
\end{verbatim}
The calling substitution for atom \texttt{p(Y)} is the sequence
$$\begin{array}{lclcclclcclclccl}
 T_1 & \longrightarrow & f(a) & & &
 T_2 & \longrightarrow & f(f(a)) & & &
 T_3 & \longrightarrow & f(f(f(a))) & & &
 \ldots
\end{array}$$
whereas the type $T \longrightarrow f(a) \ | \ f(T)$ correctly
describes such calls. However, the analysis is not able to infer such
a type.
\end{08example}

The problem in the above example is that none of the approximations
$T_i$ contains a reference to the previous approximation. This is
originated in the program fact for predicate \texttt{q/2} which
causes the loss of the reference to the previous approximation because
of the double occurrence of constant \texttt{a}.

In our analysis, termination is guaranteed by a bound on the number of
times the widening operation can be applied to a type name. A counter
is associated to each type name, so that when the bound is reached
a more restrictive widening that guarantees termination is applied. 

\section{Analysis Results}
\label{sec:results}

We have implemented analyses based on most of the widenings discussed
in this paper, including structural widening. The implementation is in
Prolog and has been incorporated to the CiaoPP
system~\cite{ciaopp-iclp99-tut},
which uses the top-down analysis algorithm of PLAI. The analysis
of~\cite{gallagher-types-iclp94}, based on regular approximations,
which uses a bottom-up algorithm, is also incorporated into the
system. This analysis uses shortening. We want to compare the top-down
and bottom-up approaches with the same widening and similar
implementation technology,\footnote{Similar in the programming
  technique. Of course, the regular approximation method is rather
  different from the method of program interpretation on an abstract
  domain: Evaluating this difference is part of the aim of the
  comparison.}
as well as the precision and efficiency,
within the same analysis framework, of the widening operators
previously discussed. 

We have used two sets of benchmark programs: the one used in the PLAI
framework and that used in the GAIA~\cite{LeCharlier94:toplas} framework.
A summary of the benchmarking follows. The analysis times in
miliseconds are shown in Table~\ref{table:bothfixp} (left). The first column
(\texttt{rul}) is for the regular approximation analysis and the other
three for the PLAI-based analyses: column \texttt{short} for
shortening, column \texttt{clash} for topological clash, and 
column \texttt{struct} for structural widening.

\newcommand{\mejor}{$^*$}

\begin{table}[htbp]
  \begin{center}
    \begin{tabular}{||l||r|r|r|r||} 
\hline \hline 
Program & \multicolumn{1}{|c|}{\texttt{rul}} & 
          \multicolumn{1}{|c|}{\texttt{short}} & 
          \multicolumn{1}{|c|}{\texttt{clash}} & 
          \multicolumn{1}{|c||}{\texttt{struct}} \\
\hline \hline 
aiakl       &  568&        469&        529&      900 \\
bid         & 1480&       2209&       2529&     4730 \\
boyer       & 3450&       3890&       4989&     9629 \\
browse      &  758&        380&        389&      539 \\
cs\_o       & 3840&       1889&       2689&     2580 \\
cs\_r       &18549&      10720&      24479&    19560 \\
disj\_r     & 4468&       1819&       6399&     2440 \\
gabriel     & 1549&       1430&       1870&     1760 \\
grammar     &  330&        160&        160&      190 \\
hanoiapp    &  620&        719&       1889&     1150 \\
kalah\_r    & 1520&         79&         79&       89 \\
mmatrix     &  310&        190&        209&      119 \\
occur       &  380&        219&        330&      289 \\
palin       &  590&        840&        980&      850 \\
pg          &  839&       2020&       2980&     3990 \\
plan        & 1138&        819&        960&     1009 \\
progeom     &  979&       1840&       2530&     3640 \\
qsort       &  310&        590&        659&      680 \\
qsortapp    &  369&       1000&       2898&     1210 \\
queens      &  329&        179&        190&      180 \\
query       &  720&        360&        370&      410 \\
serialize   &  478&        810&        969&      899 \\
witt        & 2929&       4890&       1399&     1169 \\
zebra       &  560&       3490&      14958&    12830 \\
\hline \hline 
\multicolumn{5}{c}{(excluding simplification times)}
    \end{tabular}\hfill
    \begin{tabular}{||l||r|r|r|r||} 
\hline \hline 
Program & \multicolumn{1}{|c|}{\texttt{rul}} & 
          \multicolumn{1}{|c|}{\texttt{short}} & 
          \multicolumn{1}{|c|}{\texttt{clash}} & 
          \multicolumn{1}{|c||}{\texttt{struct}} \\
\hline \hline 
aiakl       &        697&       3009&       3738&     1409 \\
bid         &       2899&      31278&      35949&    15259 \\
boyer       &      19620&     201169&     206917&    92117 \\
browse      &        987&       2848&       2987&     1698 \\
cs\_o       &      11958&      17389&      32959&     4878 \\
cs\_r       &      50760&     303430&     238788&    30169 \\
disj\_r     &       6508&      18598&      26077&     6408 \\
gabriel     &       2098&      13388&      22379&     5208 \\
grammar     &        759&       3169&       3169&     1279 \\
hanoiapp    &        840&       3988&      13738&     3378 \\
kalah\_r    &       2069&       1187&       1188&      888 \\
mmatrix     &        757&       1769&       2078&      488 \\
occur       &        530&       1647&       2628&      767 \\
palin       &        997&       8520&      11878&     2180 \\
pg          &       1349&      15380&      22870&     7370 \\
plan        &       1587&       6167&       6559&     2288 \\
progeom     &       1358&      12800&      17598&     6679 \\
qsort       &        520&       3439&       4168&     1409 \\
qsortapp    &        569&       7789&       9669&     2900 \\
queens      &        457&       1128&       1138&      429 \\
query       &       1627&      22458&      22788&    11818 \\
serialize   &        937&       8429&      11957&     2217 \\
witt        &       3438&     188419&      42699&    25709 \\
zebra       &        717&      55100&     189949&    44540 \\
\hline \hline
\multicolumn{5}{c}{(including simplification times)}
    \end{tabular}
    \caption{Timing results}
    \label{table:bothfixp}
    \label{table:bothsimp}
  \end{center}
\vspace{-2.5\baselineskip}
\end{table}

Table~\ref{table:precision} shows results in terms of precision.
The precision of \texttt{struct} is never improved by any of the
others. The improved precision of \texttt{struct} has been measured
as follows. The left subcolumns under \texttt{rul}, \texttt{short}, and
\texttt{clash} show the number of types with a more precise definition
inferred by \texttt{struct}. The right subcolumns show the number of types 
where the previous ones appear (and are thus, also, more precise).
The former are types directly inferred from program predicates; the
latter are types which are defined from the former, due to the data
flow in the program.

\begin{table}[htbp]
  \begin{center}
    \begin{tabular}{||l||r|r||r|r||r|r||r||}
\hline \hline 
Program & \multicolumn{2}{|c||}{\texttt{rul}} & 
          \multicolumn{2}{|c||}{\texttt{short}} & 
          \multicolumn{2}{|c||}{\texttt{clash}} \\
\hline \hline 
aiakl          & 1 & 1 & 1 & 1  &   &    \\  
bid            & 9 & 12& 9 & 12 &   &    \\
cs\_o          & 4 & 18& 4 & 18 & 2 & 9  \\
cs\_r          & 4 & 28& 4 & 28 & 2 & 19 \\
disj\_r        & 6 & 13& 6 & 13 &   &    \\
mmatrix        & 2 & 2 & 2 & 2  &   &    \\
occur          & 1 & 1 & 1 & 1  &   &    \\
palin          & 2 & 4 & 2 & 4  &   &    \\
pg             & 1 & 1 & 1 & 1  &   &    \\
qsort          & 1 & 1 & 1 & 1  &   &    \\
serialize      & 2 & 4 & 2 & 4  &   &    \\
zebra          & 3 & 3 & 3 & 3  & 1 & 1  \\
\hline \hline 
    \end{tabular}
    \caption{Precision results}
    \label{table:precision}
  \end{center}
\vspace{-2\baselineskip}
\end{table}

%% Also, mention the times the bound on the number of widenings is
%% reached... 

The following conclusions can be drawn from the tables. First, the
regular approximation approach seems to behave better in terms of
efficiency than the program interpretation approach, at least for the
bigger programs. This conclusion, however, has to be taken with some
care, since the current implementation of \texttt{rul} performs some
caching of the type grammars that the PLAI-based analysis does
not. % The effect of 
This should be subject of a more thorough
evaluation, which is out of the scope of this paper. The fact that it
improves in bigger programs seems to suggest that the effect of this
caching is most surely not negligible.

Regarding the analyses based on program interpretation, it can be
concluded that the better the precision the worse the efficiency:
\texttt{short} takes less than \texttt{clash}, and this one takes less
than \texttt{struct}; this one is more precise than \texttt{clash},
which is more precise than \texttt{short}. This conclusion seems
evident at first sight, but it is not: in analysis, an improvement in
precision can very well trigger an improvement in efficiency. This can
also be seen in the tables in some cases, the most significant
probably being \texttt{zebra}. % in table~\ref{table:ciaoppfixp}.
Overall, one can arguably conclude that the efficiency loss found is
not a high price in exchange for the gain in precision.
%\marginpar{discuss results!}

We have also carried out another test.
For practical purposes, the CiaoPP system includes a back-end to the
analysis that simplifies the types inferred, in the sense that
equivalent types are identified, so that they are then reduced to a
single type. This facilitates the interpretation of the output. 
It is the case that the structural widening includes certain amount of
type simplification, so that the analysis creates less different types
which are in fact equivalent. For this reason, we have included the
same tests as above, but adding now the times taken in the back-end
simplification phase. 

The times including the simplification are shown in
table~\ref{table:bothsimp}. The columns
read as before. It can be seen that in this case structural widening
outperforms all of the other analyses, except, in some cases,
\texttt{rul}. 
It also can be observed that \texttt{rul} behaves usually better than
\texttt{short} also when simplification is included. This seems to
suggest that incorporating our widening into the regular approximation
approach would probably give the best results in practice.\footnote{This,
  however, may not be trivial. It is subject for future work.}

\section{Conclusions}
\label{sec:conclusions}

We have presented a new widening operator on regular types within
an abstract interpretation-based characterization of type inference.
The idea behind it is similar to set-based
analyses~\cite{FruewirthShapiroVardiYardeni91,set-based-popl} in that
we assign and fix type names, but it is applied here with more generality.
It can be seen as a generalization of the idea of ``guessing'' the
growth of the types during analysis which is
behind~\cite{VanHentenryckCortesiLeCharlier95}. Instead of guessing,
our technique determines exactly where the type is growing.
The resulting widening operator has been presented on deterministic
regular types. However, its extension to non-deterministic regular
types should be straightforward.

Our operator is more precise than previous approaches,
but it is still efficient. This has been shown with (preliminary)
practical results. However, it does not guarantee termination. We are
currently working on the non-termination problem. A moded type domain
will help in this. The idea is to enhance abstract unification so that
it is able to identify the ``transference'' of type names from the
input to the output types, so that the names are not dropped. This
will remedy the problem of Example~\ref{ex:nontermination} and,
hopefully, allow us to prove termination of analyses with the proposed
widening operator.

Finally, this work has revealed two issues that may be worth
investigating for practical purposes: the impact on the
efficiency of analysis of the different implementation techniques
for different analysis methods,
%(e.g., regular approximations versus program interpretation),
on one hand, and of the simplification of
types, on the other hand. 

\paragraph{\bf Acknowledgements}
We would like to thank John Gallagher for very useful discussions, and
Pedro L\'{o}pez for his help with the implementation and the
availability of his library on type manipulation.
This work has been partially supported by Spanish MCYT project EDIPIA
TIC99-1151.

\end{document}